\documentclass[final]{svjour3}
\usepackage{graphicx}
\usepackage{subcaption}
\usepackage{wrapfig}
\usepackage{rotating}
\usepackage{amssymb}
\usepackage{amsmath}
\usepackage{mathptmx}
\usepackage[numbers]{natbib}

\makeatletter
\journalname{Journal of Low Temperature Physics}

\bibpunct{}{}{,}{s}{}{,}

\begin{document}

\newcommand{\hdblarrow}{H\makebox[0.9ex][l]{$\downdownarrows$}-}
\title{Identifying drivers of energy resolution variation in a multi-KID phonon-mediated detector}

\author{Karthik Ramanathan \and Taylor Aralis \and Ritoban Basu Thakur \and Bruce Bumble \and Yen-Yung Chang  \and Osmond Wen \and Sunil R. Golwala}

\date{Received: 06/08/2022}

\institute{K. Ramanathan \and T. Aralis \and R. Basu Thakur \and Y.-Y. Chang  \and O. Wen \and S. R. Golwala\\ 
Division of Physics, Mathematics, \& Astronomy, California Institute of Technology, Pasadena, CA 91125, USA \\
\email{karthikr@caltech.edu} \\
\\
B. Bumble \\
NASA Jet Propulsion Laboratory, Pasadena, CA 91107, USA}

\maketitle

\begin{abstract}

Phonon-mediated Kinetic Inductance Detectors (KIDs) on silicon substrates have demonstrated both $\mathcal{O}$(10) eV energy resolution and mm position resolution when used as particle detectors, making them strong candidates for instrumenting next generation rare-event experiments such as in looking for dark matter. Previous work has demonstrated the performance of an 80-KID array on a Si wafer, however current energy resolution measurements show a $\sim$25$\times$ difference between otherwise identical KIDs \textendash\ between 5 to 125 eV on energy absorbed by the KID. Here, we use a first principles approach and attempt to identify the drivers behind the variation. In particular, we analyze a subset of 8 KIDs using the unique approach of pulsing neighboring KIDs to generate signals in the target. We tentatively identify differences in quality factor as the likely culprit for the observed differences.


\end{abstract}

\section{KID Based Phonon Mediated Detectors}
\begin{figure}[h]
  \vspace{-10pt}
  \centering
    \includegraphics[width=0.99\linewidth, keepaspectratio]{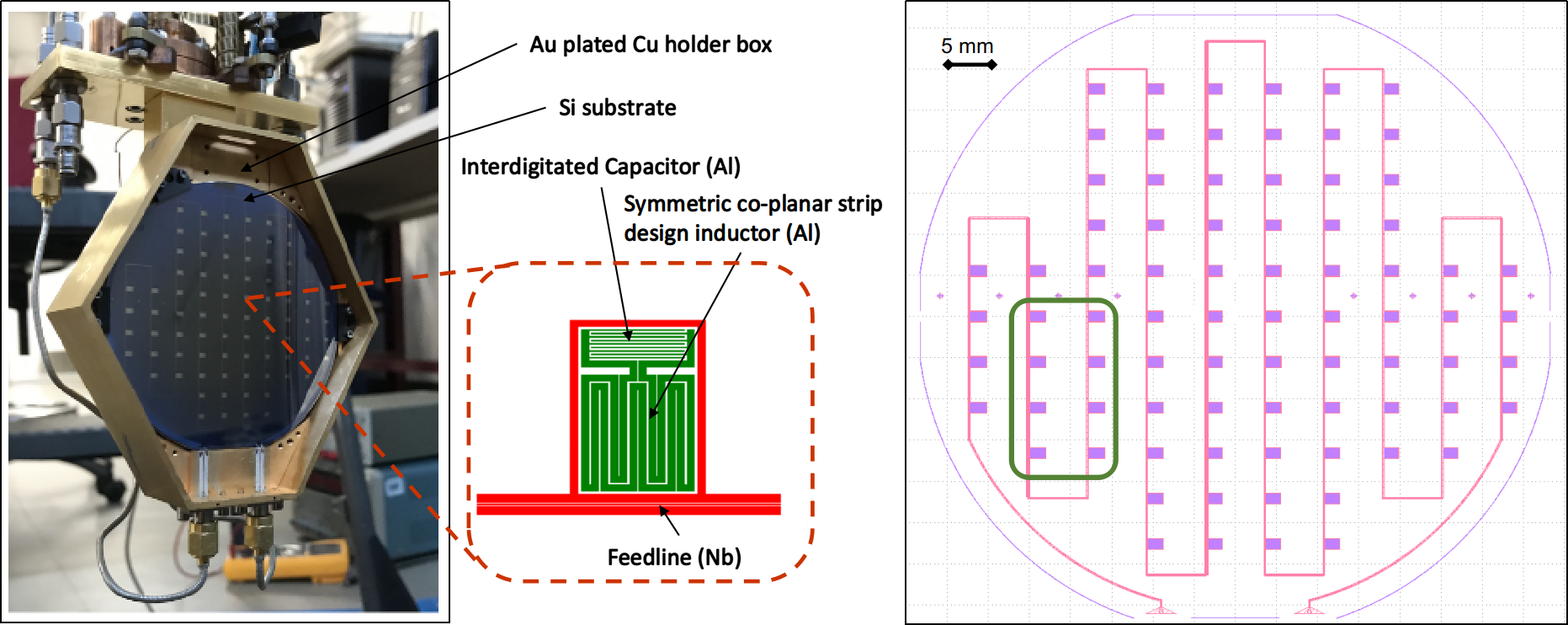}
  \caption{\textit{Left:} 80 KID device on a 75~mm diameter, 1~mm thick Si. substrate, mounted in a gold plated copper box. Cartoon zoom of a single resonator element, showing 300~nm Niobium feedline coupled to an aluminium KID composed of an interdigitated capacitor and an approx. 30-nm thick meandering symmetric coplanar inductor. \textit{Right:} Device schematic outlined with the 8 resonators considered in this analysis, as described in the text.}
  \label{fig:device}
  \vspace{-30pt}
\end{figure}

Recent dark matter detection community reports [1,2] stress the need for a roadmap to eV and sub-eV energy thresholds to probe the increasingly relevant sub-GeV$/c^2$ mass dark matter parameter space. Kinetic Inductance Detectors, as first proposed by Day et al. [3], are excellent phonon-sensing devices for these applications due to their low energy threshold, inherent multiplexability, and straightforward cryogenic RF readout. Interacting particles within the substrate bulk produce an athermal phonon population, which propagate to the surface KID film and effect a change in the quasiparticle density of the superconducting material by breaking Cooper pairs. The subsequent modified `kinetic inductance' then modulates the RF transmission properties of the KID resonator and by measuring said transmission allows one to work back through the chain to figure out details of the original energy deposit. Wen et al. [4] have demonstrated single KID resolution down to 6~eV for energy received by the phonon sensor. Previous work by Moore et al. [5] has shown how patterning 20 KID resonators on a 4~cm$^2$ silicon substrate enabled sub-mm position reconstruction and $\mathcal{O}$(100)~eV energy resolution of external radiation. Subsequent work by Chang et al. [6] led to building an 80-KID device on a 75~mm diameter $\times$ 1 mm thick Si substrate, as seen in Fig. \ref{fig:device} \textit{Left}, operated at 60~mK in an Oxford Kelvinox-25 dilution refrigerator. This prototype detector couples all KIDs to a single 300~nm wide niobium ($\Delta=$1.5~meV superconducting gap) coplanar waveguide (CPW) feedline. The capacitive and 30-nm thick inductive elements are made of aluminium ($\Delta=$200~$\mu$eV). The resonators are over-coupled, with the coupling quality factor Q$_c$ smaller than the intrinsic quality factor $Q_i$, to allow recovery of phonon rising edge information. All KIDs are identical other than small inductor length changes to separate their respective resonant frequencies $f_r$ by $\sim$5~MHz in the 3.05\textendash3.45~GHz band. KID output is fed to a 4~K noise temp. cryogenic HEMT amplifier and data is acquired using an Ettus Research SDR. As designed, the expected substrate energy resolution $\sigma_E$ is $<20$~eV across all KIDs.   

\section{Energy resolution estimation}
\begin{figure}[h]
    \vspace{-12pt}
    \centering
    \includegraphics[width=0.99\linewidth, keepaspectratio]{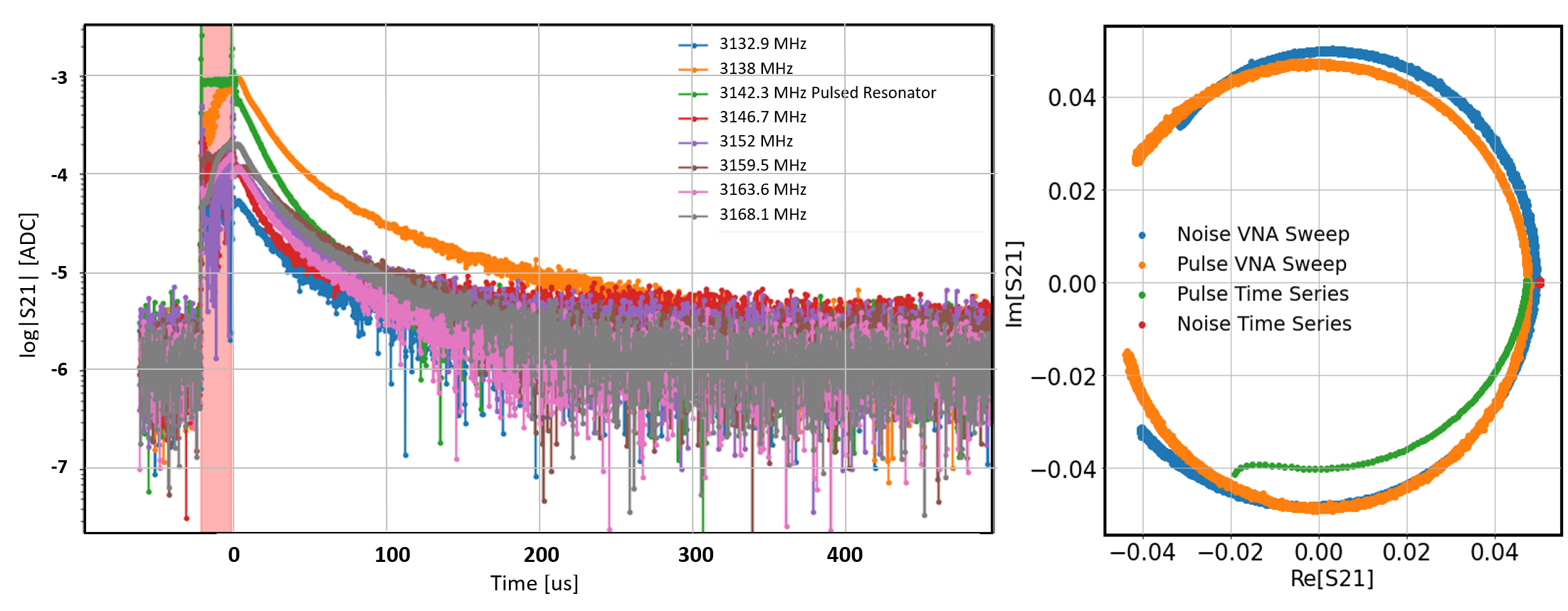}
    \caption{\textit{Left:} Driving the 3142~MHz KID (green line) with a 20~$\mu$s square pulse (pink shaded region) of approx. -30~dBm device power, results in quasiparticle production and subsequent absorption in neighboring resonators. \textit{Right:} S$_{21}$ view of different datasets taken using the 3142~MHz resonator. The pulse in this case was one received by pulsing a neighboring KID.}
    \label{fig:combined}
    \vspace{-36pt}
\end{figure}

The energy resolution can be estimated using a novel in-array technique without use of a known energy external radiation source. By pulsing a source KID with large readout power, one creates a non-equilibrium quasiparticle population within the sensor. Recombination then generates phonons that propagate out into the substrate. These are absorbed by other target KIDs, like a regular particle interaction. Fig. \ref{fig:combined} details this process, showing the response of neighboring KIDs to a 20~$\mu$s square pulse. By measuring the shift in resonance parameters and using Mattis-Bardeen (M-B) theory \textemdash\ which describes the electrodynamics of thin-film superconductivity \textemdash\ to relate the extracted parameters to the physically relevant quasiparticle density, one can then apply an optimal filter (OF) [7] to compute the resolution: 

\begin{equation}
    \sigma^2 = \int_{-\infty}^{\infty}df J(f)|\frac{\tilde{s}(f)}{J(f)}|^2 \bigg/ \bigg[ \int_{-\infty}^{\infty}df \frac{|\tilde{s}(f)|^2}{J(f)} \bigg] ^2
    \label{eq:of}
\end{equation} 
where $\tilde{s}$ is the Fourier transform of a pulse signal time stream and J is the power spectral density of a corresponding noise stream. For the analysis presented here we selected a subset of 8 resonators on the prototype device, as seen in Fig. \ref{fig:device} \textit{Left}, chosen to be both neighbors in frequency and physical space \textemdash\ downplaying position effects arising from phonon propagation across the entire wafer \textemdash\ with $f_r$ spanning from 3130\textendash3170 MHz. Pulse data was acquired by applying a series of approx. -30~dBm squares pulses of 20~$\mu$s width, in line with the phonon rise-time for these devices, to individual resonators. Resonator data and related noise time streams were then read out at -60 dBm. General analysis and processing details can be found in e.g. Ref. [8]. Device cross-talk and resonance shift overlaps are not expected as the resonators are well separated in frequency intervals with linewidths of $\Delta f/f_r \sim 0.03$. The maximum integrated pulse energy deposit into the substrate is only of $\mathcal{O}$(MeV), in line with external particle backgrounds, and is restricted to a single event per data run, negating concerns of undue wafer heating. Applying the OF framework shows large variation in measured energy resolution \textemdash\ the best performing KIDs hit the design goal of $\sigma_E=16$~eV (5~eV on what is absorbed by the sensor) while the worst performer has $\sigma_E=415$~eV (125~eV at the sensor). This variation is not explicable under amplifier-limited noise models. 

\section{Impedance mismatches and measuring quality factors}
\begin{figure}[h]
    \vspace{-6pt}
    \centering
    \includegraphics[width=0.91\linewidth, keepaspectratio]{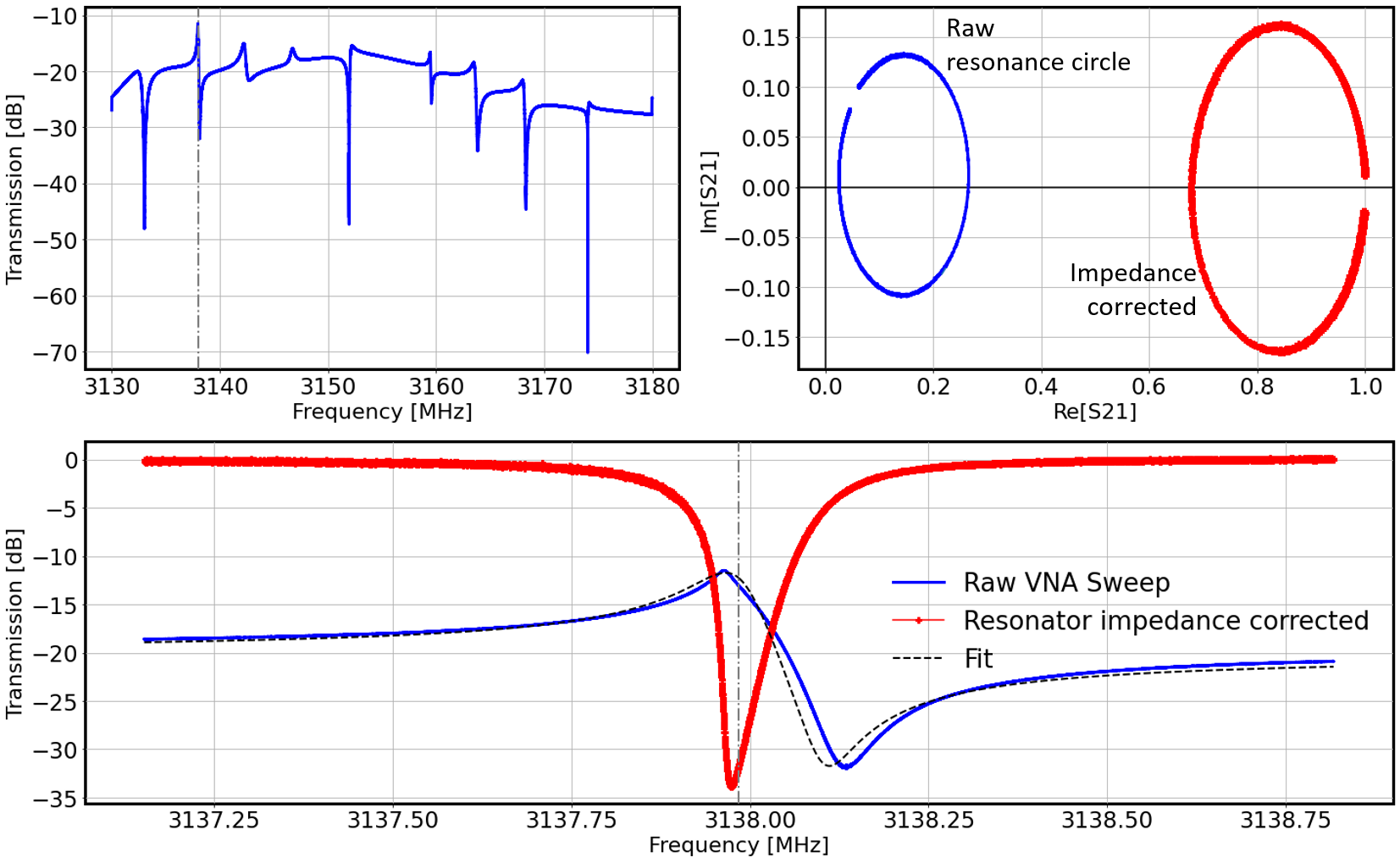}
    \caption{\textit{Top Left:} S21 Transmission across all studied resonators, showing the characteristics dips. Note the uneven overall level and upward spiking features. \textit{Top Right, Bottom:} Resonance circles and transmission spectra (\textit{legend shared}) of the 3138 MHz resonator before and after the impedance match corrections of [9], showing the effect of the $\phi$ rotation and scaling by cos($\phi$) (referenced in Eq. \ref{eq:khalil}), recovering the expected Lorentzian transmission.} 
    \label{fig:rotations}
    \vspace{-15pt}
\end{figure}

The actual transmission spectra of real devices, e.g. in Fig.~\ref{fig:rotations} \textit{Top}, shows deviations in the resonance circle from the expected transmission $S_{21}=1-(Q_r/Q_c)/(1+2jQ_rx)$ (with $x\equiv(f-f_r)/f_r$) at resonance for total quality factor $Q_r$, where $Q_r^{-1}=Q_c^{-1}+Q_i^{-1}$. These lead to an asymmetric transmission line shape even at low-power. Khalil et al. [9] attribute this to an ``impedance mismatch" between the input and output lines of the resonator and argue that simple technical corrections, such as rotating the resonance circle without accounting for transmission changes, can lead to divergent over-estimates of $Q_{i}$ for both low-$Q$ devices and rotation angles close to $\pi$ ($\sim$30\% in some of the resonators considered here). They quantify the mismatch by introducing an imaginary component to $Q_c$, parametrized by a rotation angle $\phi$, leading to a modified description of the transmission:
\begin{equation}
    S_{21}(f) = a e^{-2\pi j f \tau} \big[ 1 - [ (Q_r/Q_c \rm{cos}\phi) e^{j\phi} \big/ (1+2jQ_r x) ] \big]
    \label{eq:khalil}
\end{equation}
which includes a complex feedline attenuation term $a$ and a feedline delay term $\tau$.

Measuring individual terms in Eq.~\ref{eq:khalil} is nominally achieved by fitting to the raw transmission spectrum, as seen in Fig. \ref{fig:rotations} \textit{Bottom}, with the caveat that the fit is potentially degenerate in the free terms of \{$Q_r$,$Q_c$,$a$,$\tau$,$\phi$\}. Additionally, one notes the imperfect fit, perhaps indicating that the impedance-mismatch model is an incomplete description of resonator behavior. One then uses the M-B relations $\delta (1/Q_i) \approx \alpha \kappa_1 n_{qp}$ and $\delta f_r/f_r \approx -\alpha \kappa_2 n_{qp}/2$ [10], where $\alpha$ is the material and geometry based fraction of the total inductance of the resonators due to kinetic inductance, and $\kappa_{1,2}$ are temperature sweep derived parameters to help move to a quasiparticle number $n_{qp}$ basis from which we can compute the ``fitted" resolution $\sigma_{\rm{fit}}$ via Eq. \ref{eq:of} as previously disclosed. However, starting with Eq. \ref{eq:of}, we can generally break up the energy resolution (expressed as a $n_{qp}$ number density) into a linear form with constituents:

\begin{equation}
    \sigma^{-2} = \bigg( \sum{\frac{|\tilde{s}|^2}{J}} \bigg) \cdot a^2 \cdot R_Q^2 \cdot (\alpha \kappa_{1,2})^2
    \label{eq:res_np}
\end{equation}
where we have bundled up the quality-factors into a single term $R_Q$ which in the language of Eq. \ref{eq:khalil} would be $\equiv Q_r^2/(Q_c\cdot \rm{cos}\phi)$. Identifying the source of the variation in the energy resolution now becomes a matter of investigating each of the four terms in the RHS of Eq. \ref{eq:res_np}. The first term is computed from the raw noise and pulse time stream data in the electronics basis and encodes the detector noise response. The complex attenuation is frequency dependent and is linked to the overall transmission as seen in Fig. \ref{fig:rotations} \textit{Top Left}. The M-B parameters have been measured across all resonators and found to be within 1\%. Crucially $R_Q$ can be extracted from transmission data without disambiguating \{$Q_r$,$Q_c$,$\phi$\} \textemdash\ by identifying the freq. direction of the resonance circle, knowing the readout frequency of each data point, estimating $x \approx \delta f_r/f_r$, and recasting the change in transmission as follows:
\begin{equation}
    \delta S_{21} \sim R_Q(\delta\frac{1}{Q_i}-2j\frac{\delta f_r}{f_r})  
    \implies R_Q \approx \frac{1}{2}\frac{\delta S_{21}}{\delta x}
    \label{eq:empirical}
\end{equation}

\section{Device Performance and Discussion}

\begin{figure}[h]
    \vspace{-12pt}
    \centering
    \includegraphics[width=0.99\linewidth]{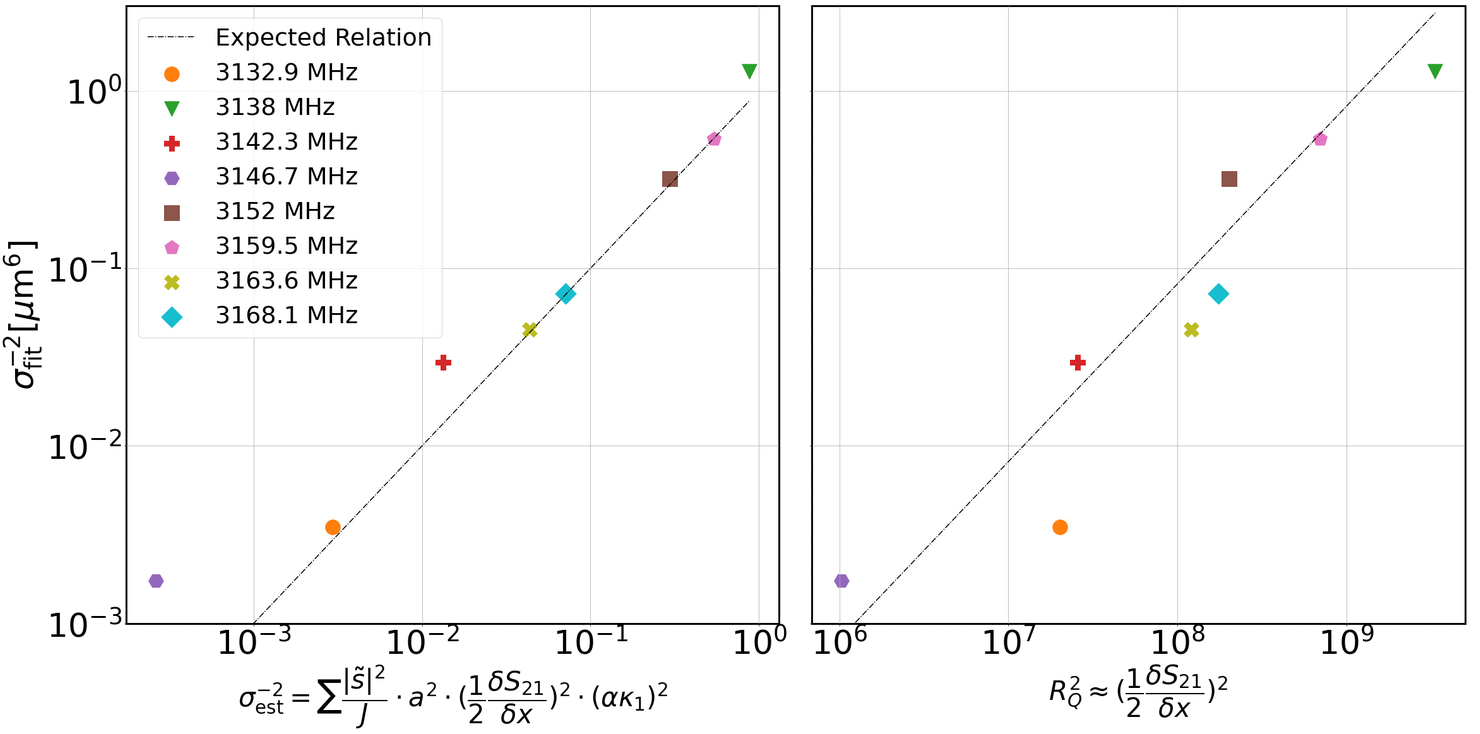}
    \caption{Expected energy resolution from measurement in the $\kappa_1$ direction at -60~dBm readout power, expressed in quasiparticle units, for the various studied resonators. \textit{Left:} $\sigma_{\rm{est}}$ as computed from its constituents and compared to the fitted resolution, showing good agreement between both methods for the majority of resonators. \textit{Right:} Resolution as related to the empirically measured $R_Q$. Differences in this factor appear to drive the variation in resolution.}
    \label{fig:resprediction}
    \vspace{-15pt}
\end{figure}

Computing this empirical resolution $\sigma_{est}$ through Eqs. \ref{eq:res_np} \& \ref{eq:empirical} for each resonator, we arrive at Fig. \ref{fig:resprediction} which shows the example of the $\kappa_1$ direction energy resolution. The \textit{Left} plot shows the expected unity relationship between the modeled $\sigma_{\rm{fit}}^{-2}$ and the constituent derived $\sigma_{\rm{est}}^{-2}$. Statistical errors are only of order 10\% with deviations between both series likely due to systematics in their respective reconstruction processes. More importantly, the \textit{Right} plot shows how the resolution appears to be entirely driven by differences in the quality factor ratio $R_Q$ versus alternative explanations like varied noise performance between resonators. As a side note, we can convert from the quasiparticle basis to a substrate deposited energy resolution using $\sigma_{\rm{eV}} = \sigma_{\rm{nqp}} \cdot \Delta \cdot V / \eta_{\rm{ph}}$, with a KID volume $V$=3$\cdot$10$^4$~$\mu$m$^3$ and a phonon to quasiparticle conversion efficiency at the substrate-detector interface via $\eta_{\rm{ph}}\approx0.3$.

\begin{figure}[h!]
    \vspace{-6pt}
    \centering
    \includegraphics[width=0.99\linewidth]{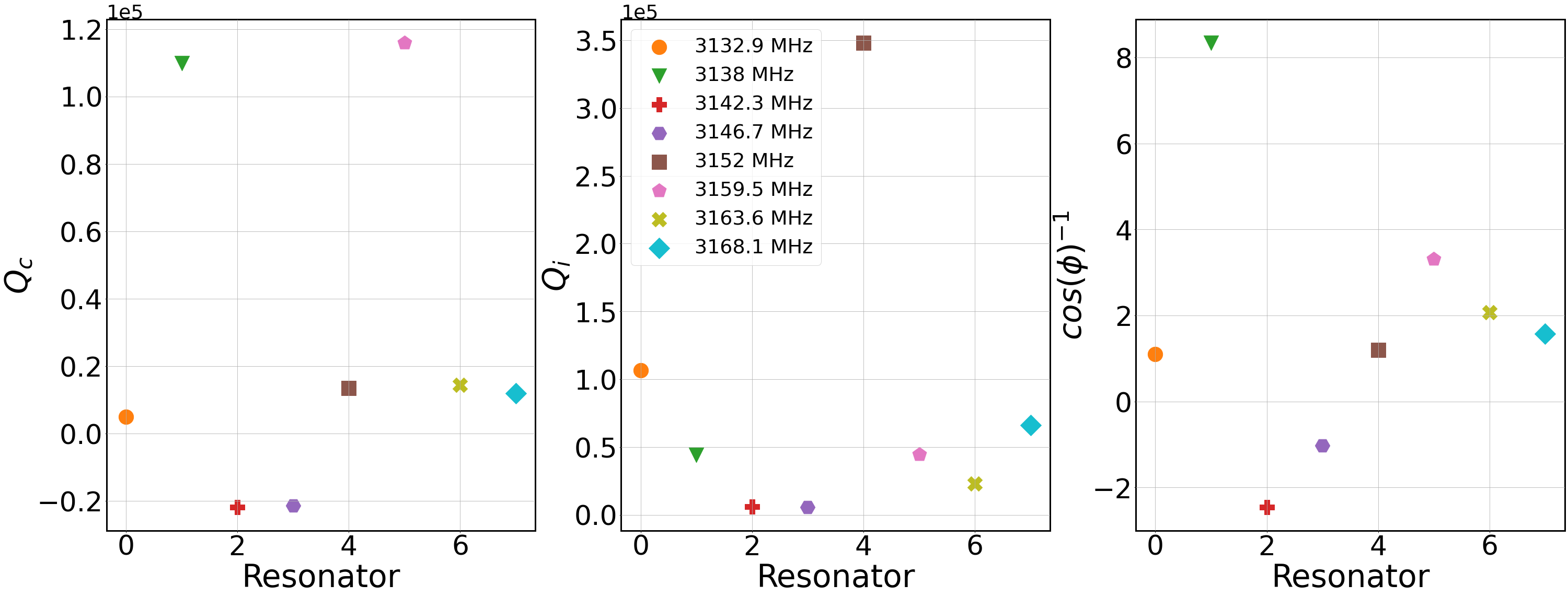}
    \caption{Internal quality factor (\textit{Left}), coupling quality factor (\textit{Center}), and impedance mismatch $\phi$ (\textit{Right}), for all 8 studied resonators, showing wide variation between devices. Statistical uncertainties are suppressed but are at the percent level.}
    \label{fig:qiqc}
    \vspace{-15pt}
\end{figure}

\begin{figure}[h]
    \vspace{-6pt}
    \centering
    \includegraphics[width=0.99\linewidth, keepaspectratio]{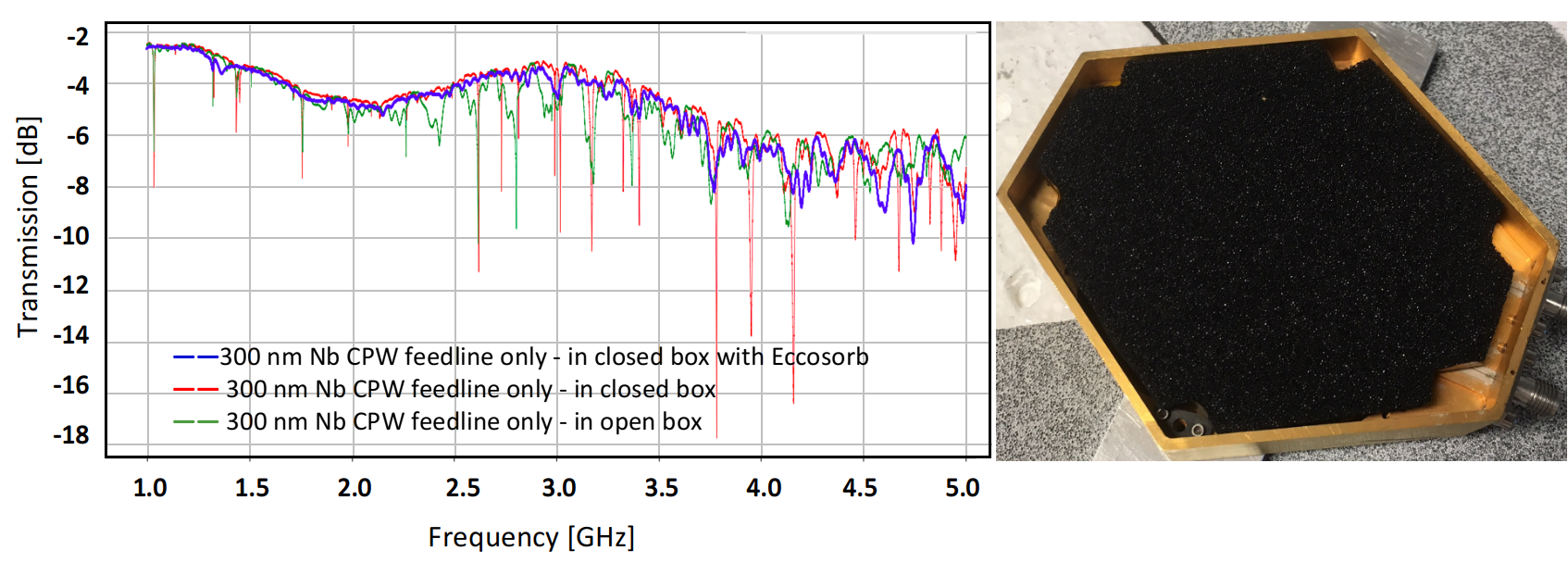}
    \caption{\textit{Left:} 4K measured S$_{21}$ transmission of a 300~nm wide Nb CPW feedline for different configurations of the Cu box. \textit{Right:} Box with Eccosorb cutout layer held on by clamps.}
    \label{fig:boxmode}
    \vspace{-12pt}
\end{figure}

Fig.~\ref{fig:qiqc} shows the extracted components $Q_c$, $Q_i$, and $\phi$ of $R_Q$, mindful that individual quantities are perhaps degenerate. It does not reveal a single culprit for the resolution variation. The devices that show the greatest deviation from the pack, e.g. the 3138 MHz resonator, are the same ones that show the largest difference in resolution in Fig.~\ref{fig:resprediction}. We note certain unexpected negative $Q_c$, though offset by a corresponding negative cos($\phi$) term. We can only conclude that the overall combination of terms has an effect, and interpreting the change as arising from specific models (e.g. impedance matching, resonator differences) is thus not straightforward.
One hypothesis for the observed behavior is the presence of box modes, i.e. EM couplings between the device and its metallic holder box, sourcing the changing performance in a more complicated way than the impedance mismatch model. Some support for this hypothesis was established by measuring the transmission for a feedline only device, as seen in Fig. \ref{fig:boxmode}, in different configurations. In the default closed box (red line), the numerous spectral line features are indicative of these box modes. Other spectral features are apparent with the device lid off (green line), but these modes are completely removed after applying a thin $\sim$mm layer of Eccosorb dielectric foam absorber (blue line), implying that the transmission diagram in Fig. \ref{fig:rotations} \textit{Top Left} can be cleaned up with an absorber. However this can potentially degrade resonator quality factors, because the Eccosorb can remain at an elevated temp. and act like a blackbody load on the device. 

\section{Conclusion and Future Work}
In this letter we have highlighted an ongoing concern in deploying large scale KID arrays, in that the energy resolution across devices is inconsistent even with identical designs. We used an empirical method to extract the quality factors for the resonators, were able to reconstruct the measured energy resolution from its constituent components, and identified the variation as likely arising from changing quality factor ratios. We briefly discussed a possible source of this variation as coming from box modes but accurately pinning it down will require further experiment and simulation work. Eliminating these differences between KIDs will be a necessary step towards deploying detectors with $\mathcal{O}$(100) KIDs and realizing the promise of $\mathcal{O}$(10)~eV energy resolution necessary for future dark matter searches [1,2].

\begin{acknowledgements}
We acknowledge the support of the following institutions and grants: NASA, NSTGRO 80NSSC20K1223; Department of Energy, DE-SC0011925F; Fermilab, LDRD Subcontract 672112
\end{acknowledgements}

\end{document}